\documentclass[twocolumn,pre]{revtex4}

\usepackage{amsfonts,amsmath,amssymb}
\usepackage[dvips]{graphicx,color}
\input epsf

\newcommand{\dir}{Figs}

\newcommand{\Real}{\mathop{\rm Re}\nolimits}

\begin{document}
\title{Influence of sequence correlations on the adsorption 
      of random copolymers onto homogeneous planar surfaces}

\author{Alexey Polotsky}
\email{polotsky@Physik.Uni-Bielefeld.de}
\affiliation{Fakult\"at f\"ur Physik, Universit\"at Bielefeld,
Universit\"atsstra{\ss}e 25, D-33615 Bielefeld, Germany}

\author{Friederike Schmid}
\affiliation{Fakult\"at f\"ur Physik, Universit\"at Bielefeld,
Universit\"atsstra{\ss}e 25, D-33615 Bielefeld, Germany}

\author{Andreas Degenhard}
\affiliation{Fakult\"at f\"ur Physik, Universit\"at Bielefeld,
Universit\"atsstra{\ss}e 25, D-33615 Bielefeld, Germany}

\begin{abstract}
Using a reference system approach, 
we develop an analytical theory for the adsorption of random heteropolymers 
with exponentially decaying and/or oscillating sequence correlations on
planar homogeneous surfaces. We obtain a simple equation for the 
adsorption-desorption transition line. This result as well as the
validity of the reference system approach is tested by a comparison
with numerical lattice calculations.
\end{abstract}

\maketitle

\section{Introduction}

\label{Introduction} 

Random heteropolymers (RHPs) currently attract much attention and 
have been studied intensely in the last two decades. The reason
for the growing interest in them is twofold. On the one hand, 
they can be used to create new polymeric materials, or to modify and 
improve surface and interface properties. On the other hand, RHPs serve 
as simple model systems for biologically relevant heterogeneous 
macromolecules, such as proteins and nucleic acids. Studying the physics 
of RHPs helps to understand generic properties of these macromolecules 
and phenomena which play an important role in living organisms,
such as protein folding, denaturation/renaturation of DNA, 
secondary structure formation by RNA, etc.


In the present paper, we consider RHP adsorption on homogeneous and 
impenetrable (solid) surfaces. This has technological relevance in the
context of adhesion (glue) and surface design. Studying RHP adsorption 
can also be regarded as a (very) first step towards understanding some 
elementary principles of molecular recognition.

For entirely random RHPs, the problem was investigated theoretically 
by a number of authors.
Joanny~\cite{Joanny} considered the adsorption of polyampholyte chains 
on a planar solid surface. In the case that the electrostatic interactions 
are strongly screened and thus short-ranged, he treated the polyampholyte 
chain as an ideal RHP with no intersegment interactions, and solved the 
problem using the replica trick and the Hartree approximation. 
A replica symmetric mean field theory that takes into account intersegment
interactions has been developed by Gutman and 
Chakraborty~\cite{Gutman1, Gutman2}, to describe RHPs interacting 
with one~\cite{Gutman2} or two parallel~\cite{Gutman1} planar surfaces. 
They analyzed the interfacial structure (as expressed through the density 
and the order parameter profiles) as well as the adsorption-desorption 
phase diagram.  
Stepanow and Chudnovsky \cite{Stepanow} combined the replica trick
with the Greens function technique to study the localization of a RHP
onto a homogeneous surface, and suggested a phase diagram for the
localization-delocalization transition. 
The system has been also studied by Monte-Carlo 
simulations~\cite{Sumithra,Moghaddam}. Sumithra and Baumgaertner~\cite{Sumithra} 
have complemented their simulations by a scaling analysis of the conformational 
and thermodynamic characteristics of the system.

In all of these works the RHPs were taken to be of Bernoullian type,
i.~e., the monomer sequences had no correlations. Sequence 
correlations should, of course, affect the adsorption properties of RHPs. 
van Lent and Scheutjens~\cite{vanLent} have formulated a self-consistent 
field theory for random copolymers with correlated sequences in the case
of {\em annealed} disorder, i.~e., the copolymer sequences can ``adjust''
to their local environment. They used it to study the adsorption
of random copolymers from solution within a lattice model.
Zheligovskaya et al.~\cite{Zheligovskaya} have explored the
idea of conformation-dependent sequence design and compared the
adsorption properties of AB-copolymers with special ``adsorption-tuned'' 
primary structures (adsorption-tuned copolymers, ATC) with those of truly 
random copolymers. Monte-Carlo simulations revealed that some specific features 
of the ATC primary structure promote the adsorption of ATC chains, compared 
to random chains.
Chakraborty and coworkers \cite{Chakraborty1,Chakraborty_review} have
demonstrated the importance of sequence correlations in a number of
studies on recognition between random copolymers and chemically 
heterogeneous surfaces.

Whereas these studies were mostly numerical, Denesyuk and 
Erukhimovich~\cite{Denesyuk} have recently presented an approximate 
analytical solution of a closely related problem, the adsorption of 
RHPs with sequence correlations on {\em penetrable} homogeneous substrates 
(e.~g., interfaces between two selective solvents). Their study is based 
on a reference system approach originally due to Chen~\cite{Chen}. 
The sequence correlations were described by a correlator introduced by 
de Gennes \cite{deGennes78} that covers both exponentially decaying and 
oscillating correlations. Denesyuk an Erukhimovich showed that chains with 
both types of correlation exhibit a continuous adsorption transition, 
and constructed the corresponding adsorption-desorption phase diagrams.

In the present paper we extend this work to study the adsorption of  
RHPs with correlated monomer sequences onto a homogeneous {\em impenetrable}
planar surface. We use the reference system approach~\cite{Chen} and
follow closely the ideas developed in Ref. \onlinecite{Denesyuk}. 
This will allow us to derive an explicit and surprisingly simple
analytical expression for the adsorption transition point. 
To test our prediction, we also performed numerical calculations on the 
lattice, and compared the results to the premises and the result 
of the analytical theory.

The rest of the paper is organized as follows: 
The analytical theory is developed in Sec.~\ref{sec:ana}.
We begin with defining the model in Sec.~\ref{sec:model_ana},
and discuss the theoretical approach in the next three 
subsections, Sec.~\ref{sec:approach}, Sec.~\ref{sec:reference},
and Sec.~\ref{sec:transition}. The final result is presented
in Sec.~\ref{sec:result_ana}. Then, we describe the numerical
calculations in Sec.~\ref{sec:num}. The numerical calculations
are compared to the analytical prediction in Sec.~\ref{sec:comparison}.
We summarize and conclude in Sec.~\ref{sec:conclusions}.

\section{Analytical Approach}
\label{sec:ana}
\subsection{Definition of the Model}
\label{sec:model_ana}

We consider an ensemble of single heteropolymer Gaussian chains, each 
consisting of $N$ monomer units, near an impenetrable planar surface. 
Monomers only interact with the surface, not with each other.
The strength and the sign of the monomer-surface interaction varies
from monomer to monomer and is characterized by an affinity parameter $\xi$. 
A particular heteropolymer realization is described by the sequence 
$\xi (n)$, where $n$ indicates the monomer number ($n=1,2,..., N$). 
It is, however, convenient to consider $n$ as a continuous rather than a
discrete variable. The Hamiltonian of the system can then be
represented as follows:
\begin{equation}
\label{Hamiltonian} \frac{H}{kT}=\frac{3}{2a^2}\int_0^N
dn\left(\frac{\partial z}{\partial n}\right)^2-\beta\int_0^N
dn \, V(z(n)) \, \xi(n),
\end{equation}
where the $z$ direction is perpendicular to the surface and
we have integrated out the \textit{x} and \textit{y} coordinates.
Here $\beta=1/kT$ is the Boltzmann factor as usual,
$a$ is the Kuhn segment length of the chain, 
and $V(z)$ gives the shape of the monomer-surface interaction potential
($\int V(z) \: dz = 1$), which is taken to be the same for all monomers. 
More specifically, we use a pseudopotential -- a Dirac well shifted by
the small but finite distance $z_0$ from the impenetrable surface 
\begin{equation}
\label{potential} V(z)=\left\{
\begin{array}{lc}
\delta(z-z_0) & \mbox{if } z>0 \\
+ \infty &  \mbox{if } z<0
\end{array}
\right.
\end{equation}
The parameter $z_0 > 0$ corresponds to the range of the surface potential.
It has been introduced mainly for technical reasons, which will
become apparent further below. In the end
(Sec. \ref{sec:result_ana}), we will take the limit $z_0 \to 0$.

The sequences $\xi(n)$ are randomly distributed according to a 
Gaussian probability function
\begin{widetext}
\begin{equation}
\label{seq_prob} P\{\xi(n)\}=
\frac{\exp\left[ -\frac{1}{2}\int_0^N dn_1 \int_0^N dn_2
(\xi(n_1)-\xi_0)c^{-1}(n_2-n_1)(\xi(n_2)-\xi_0)\right]}{\int
\mathcal{D}\{\xi(n)\}\exp\left[ -\frac{1}{2}\int_0^N dn_1 \int_0^N
dn_2 (\xi(n_1)-\xi_0)c^{-1}(n_2-n_1)(\xi(n_2)-\xi_0)\right]}.
\end{equation}
\end{widetext}
Here $\xi_0=\langle\xi\rangle$ is the mean affinity and $c(n_1, n_2)$
describes the sequence correlations. The latter is taken to have the
special form suggested by de Gennes \cite{deGennes79},
\begin{eqnarray}
\label{corrf} c(n_1, n_2) &\equiv&
\left\langle(\xi(n_1)-\xi_0)(\xi(n_2)-\xi_0)\right\rangle
\nonumber\\
&=&c(|n_2-n_1|)=\Delta^2 \Real \: e^{-\Gamma|n_2-n_1|},
\end{eqnarray}
where $\Real$ denotes the real part of a complex number. This allows us, 
by proper choice of the parameter $\Gamma$, to cover both exponentially 
decaying (real $\Gamma$) or oscillating (purely imaginary $\Gamma$) 
correlations. The expression $\int \mathcal{D}\{\xi(n)\}$ denotes functional 
integration over all possible sequences (integration in sequence space), 
and $c^{-1}(n)$ is the inverse function of $c(n)$, defined by 
$\int \! dn' \, c^{-1}(n-n')c(n')=\delta(n)$.

We will calculate the {\em quenched} average over all sequences, 
i.~e., the free energy of the system is the disorder average 
of the logarithm of the conformational statistical sum, 
$F=\langle \log Z_{\mbox{\tiny conf}} \rangle_{\xi(n)}$. For infinitely
long chains, the adsorption transition in this model is
encountered at $F=0$.

\subsection{The Reference System Approach}
\label{sec:approach}

Having defined the model, we now turn to describing the analytical approach.
To perform the quenched average we exploit the well-known replica trick
\begin{equation}
\langle \log Z \rangle_{\xi(n)} = 
\lim_{m \to 0} \langle \frac{Z^m - 1}{m} \rangle_{\xi(n)}.
\end{equation}
This reduces the problem of evaluating the disorder average 
over a logarithm, $\langle \log Z \rangle$, 
to the much more feasible task of evaluating the average partition 
function for a set of $m$ identical heteropolymer chains ($m$ replicas) 
$\langle Z^m \rangle$. 
Inserting Eqs.~(\ref{seq_prob}) - (\ref{corrf}) and using properties
of the Gaussian distribution, we obtain
\begin{widetext}
\begin{equation}
\begin{split}
\label{Z_repl_averaged} 
\langle Z^m \rangle_{\xi(n)} = & \int
\prod_{\alpha=1}^{m}\mathcal{D}\{z_{\alpha}(n)\} 
\exp\left[ -\frac{3}{2a^2}\sum_{\alpha=1}^{m} \int_0^N
dn\left(\frac{\partial z_{\alpha}}{\partial n}\right)^2 
+ \beta \xi_0 \int_0^N dn
\sum_{\alpha=1}^{m} V(z_{\alpha}(n)) \right] \\
& \times \exp \left[ \frac{\beta^2}{2}\int_0^N dn_1 \int_0^N
dn_2\sum_{\alpha=1}^{m} V(z_{\alpha}(n_1))\sum_{\gamma=1}^{m}
V(z_{\gamma}(n_2)) c(n_2-n_1)\right],
\end{split}
\end{equation}
\end{widetext}
where $\alpha$ is the replica index and $\int \mathcal{D}\{z(n)\}$
denotes the integration over all possible trajectories $z(n)$ of the
chain.

One can easily see that the problematic term in this expression
is the last factor which couples different replicas. To simplify
this term we introduce a reference homopolymer system~\cite{Chen} 
whose conformational properties are reasonably close to those of the 
original heteropolymer. It is natural to choose as reference system 
a single homopolymer chain with uniform monomer affinity $\xi_0 + v_0$,
which is exposed to a surface potential $V(z)$ of the same shape
(\ref{potential}) as the original RHP. The parameter $v_0$ is a fit 
parameter which can be adjusted such that the reference system is as close 
as possible to the original system. It is nonzero and positive, because 
the effective surface affinity of heteropolymers is generally larger 
than that of homopolymers with the same charge. For example, even 
globally neutral random heteropolymers with $\xi_0 = 0$ can adsorb 
onto a surface, whereas neutral homopolymers obviously remain desorbed.

The Hamiltonian of the reference system has the following form:
\begin{equation}
\label{Hamiltonian_ref} \frac{H_0}{kT}=\frac{3}{2a^2}\int_0^N
dn\left(\frac{\partial z}{\partial n}\right)^2-\beta(\xi_0 +
v_0)\int_0^N \, dn \, V(z(n))
\end{equation}

The difference between the free energies of the original and the
reference system is given by
\begin{equation}
-\beta (F-F_0)=\lim_{m \to 0} \frac{Z^m-Z_0^m}{m}
\end{equation}
where $Z_0^m$ is the partition function of the $m$ times
replicated reference system:
\begin{widetext}
\begin{equation}
Z_0^m= 
\int\prod_{\alpha=1}^{m}\mathcal{D}\{z_{\alpha}(n)\} 
\exp\left[ -\frac{3}{2a^2}\sum_{\alpha=1}^{m} \int_0^N
dn\left(\frac{\partial z_{\alpha}}{\partial n}\right)^2+\beta
(\xi_0+v_0) \int_0^N dn \, \sum_{\alpha=1}^{m}\ V(z_{\alpha}(n))
\right]
\end{equation}
Thus one has
\begin{equation}
\label{Z_diff}
\begin{split}
 Z^m-  Z_0^m = & \int
\prod_{\alpha=1}^{m}\mathcal{D}\{z_{\alpha}(n)\} 
\times \exp\left[ -\frac{3}{2a^2}\sum_{\alpha=1}^{m} \int_0^N
dn\left(\frac{\partial z_{\alpha}}{\partial n}\right)^2+ \beta
(\xi_0+v_0) \int_0^N dn \, \sum_{\alpha=1}^{m}\ V(z_{\alpha}(n))
\right]\\
\times &\left\{ \exp \left[ \frac{\beta^2}{2}\int_0^N dn_1
\int_0^N dn_2\sum_{\alpha=1}^{m}
V(z_{\alpha}(n_1))\sum_{\gamma=1}^{m}
V(z_{\gamma}(n_2)) c(n_2-n_1) 
-\beta v_0 \int_0^N dn \, \sum_{\alpha=1}^{m}\
V(z_{\alpha}(n)) \right] -1 \right\}
\end{split}
\end{equation}
\end{widetext}
The first exponential in this expression corresponds to the product of 
the $m$ independent propagators ${\cal P}(\xi_0, v_0, z_{\alpha}(n))$
for the $m$ chains in the replicated reference system, which give
the statistical weight of the particular trajectories $z_{\alpha}(n)$. 
Since the original and the reference systems are assumed 
to be close to each other, we expand the exponent in the second term: 
$e^A - 1 \simeq A$.  Here $A$ is the difference between Hamiltonians of 
replicated original and the reference systems $-\beta(H^m-H_0^m)$. 
This means that we approximate the difference between the free energies 
of the original and the reference system $\Delta F \equiv F_0 - F$ by its 
first cumulant $\Delta F_1$. The expansion gives
\begin{widetext}
\begin{equation}
\label{Z_diff_appr}
\begin{split}
Z^m-Z_0^m \simeq &
\int \prod_{\alpha=1}^{m}
\Big[
\mathcal{D}\{z_{\alpha}(n)\} {\cal P}(\xi_0, v_0, z_{\alpha}(n))\Big]
\\
& \times \left[ \frac{\beta^2}{2}\int_0^N dn_1 \int_0^N
dn_2\sum_{\alpha=1}^{m} V(z_{\alpha}(n_1))\sum_{\gamma=1}^{m}
V(z_{\gamma}(n_2)) c(n_2-n_1) 
-\beta v_0 \int_0^N dn \, \sum_{\alpha=1}^{m}\
V(z_{\alpha}(n)) \right].
\end{split}
\end{equation}
\end{widetext}
The second integral in Eq.~(\ref{Z_diff_appr}) is equal to
\begin{equation}
\label{I2}
\begin{split}
 I_2 & = \int
\prod_{\alpha=1}^{m}
\Big[ 
\mathcal{D}\{z_{\alpha}(n)\} {\cal P} (\xi_0, v_0, z_{\alpha}(n)) 
\Big]
\\ & \qquad \times \: \beta v_0 \int_0^N dn \, \sum_{\alpha=1}^{m}
V(z_{\alpha}(n)) \\
& = \beta v_0 N m \overline{V(z(n))} Z_0^m,
\end{split}
\end{equation}
where $\overline{\cdots}$ denotes the average with respect to the
reference system:
\begin{equation}
\label{average_def} \overline{X}= \frac{\int \mathcal{D}\{z(n)\}
G(\xi_0, v_0, z(n)) X(z(n))}{\int \mathcal{D}\{z(n)\} G(\xi_0,
v_0, z(n))}
\end{equation}
Next we consider the first integral
\begin{equation}
\label{I1}
\begin{split}
I_1 = 
& 
\int \prod_{\alpha=1}^{m}
\Big[ \mathcal{D}\{z_{\alpha}(n)\} {\cal P}(\xi_0, v_0, z_{\alpha}(n))  \Big]
\\
& \times 
\frac{\beta^2}{2} \int_0^N dn_1 \int_0^N dn_2
\sum_{\alpha=1}^{m} V(z_{\alpha}(n_1)) 
\\ & \times
\sum_{\gamma=1}^{m}
V(z_{\gamma}(n_2)) c(n_2-n_1).
\end{split}
\end{equation}
The product of the two sums in Eq.~(\ref{I1}) gives
\begin{eqnarray}
\label{double_sum}
\lefteqn{
\sum_{\alpha=1}^{m}
V(z_{\alpha}(n_1)) \sum_{\gamma=1}^{m}
V(z_{\gamma}(n_2))
} \qquad \\ &&
=  \sum_{\alpha=1}^{m} V(z_{\alpha}(n_1))V(z_{\alpha}(n_2)) 
\nonumber \\ & &
+ 
\sum_{\alpha=1}^{m} \sum_{\gamma=1 \atop \gamma \neq \alpha}^{m}
V(z_{\alpha}(n_1)) V(z_{\gamma}(n_2)),
\nonumber
\end{eqnarray}
where we have separated contributions from the same and different
replicas. This leads to the following result for $I_1$:
\begin{eqnarray}
\label{I1_2prime_final}
\lefteqn{
I_1 = \beta^2 m \, Z_0^m \int_0^N dk \, c(k) \, (N-k) \, 
} \qquad \\ && \nonumber
\Big\{
(m-1) \overline{V(z_{\alpha}(n))V(z_{\gamma}(n+k))}\Big|_{\alpha \ne \gamma}
\\ && \nonumber
+ \overline{V(z_{\alpha}(n))V(z_{\alpha}(n+k))}
\Big\}.
\end{eqnarray}
Finally, collecting all terms, taking the limit $m \to 0$, 
dividing by $N\gg 1$, and using $c(k) k/N \gg 1$ for all $k$,
we obtain for the first cumulant 
$\Delta F_1$
\begin{equation}
\label{first_cumulant}
\begin{split}
\frac{\beta \Delta F_1}{N} = 
& 
\, \beta^2
\int_0^N dk \, c(k) \, \overline{V(z_{\alpha}(n))V(z_{\alpha}(n+k))} 
\\ & 
- \beta^2 \int_0^N dk \, c(k) \, 
\overline{V(z_{\alpha}(n))V(z_{\gamma}(n+k))}\Big|_{\alpha \ne \gamma}
\\ & 
- \beta v_0 \overline{V(z_{\alpha}(n))}.
\end{split}
\end{equation}
The next step is to find the Greens function of the reference
system, which is needed for calculating the averages in
Eq.~(\ref{first_cumulant}).

\subsection{The Reference System}
\label{sec:reference}

As defined above, the reference system is a homopolymer chain with
the affinity parameter $w_0=\xi_0+v_0$, which propagates in the
half space $z > 0$ subject to the potential $V(z)=\delta(z-z_0)$. 
The Greens function of the polymer chain in the reference system 
satisfies the differential equation~\cite{DoiEdwards, deGennes_book} 
\begin{equation}
\label{diff_eq} 
\frac{\partial G}{\partial N}=
\frac{a^2}{6}\frac{\partial^2 G}{\partial z^2}
+\beta w_0 \delta(z-z_0) G(z,z';N)
\end{equation}
with the initial condition 
\begin{equation}
\label{initial_cond} G(z,z';0)=\delta(z-z')
\end{equation}
and the boundary conditions
\begin{equation}
\label{boundary_cond} 
\begin{array}{rcl}
G(z,z';N)|_{z \to \infty} & = & 0 \\
G(z,z';N)|_{z=0} &= & 0
\end{array}.
\end{equation}
Eq.~(\ref{diff_eq}) has the structure of a Schr\"odinger
equation and can be solved analogously by a separation of
variables. The solution is an expansion in eigenfunctions of 
the ``Hamilton operator'',
\begin{equation}
\label{expansion} G(z,z';N)=\sum_i e^{-\varepsilon_i
N}\psi_i(z)\psi_i^*(z'),
\end{equation}
where $\varepsilon_i$ and $\psi_i(z)$ are eigenvalues and
eigenfunctions of the stationary equation
\begin{equation}
\label{stat_diff_eq} \frac{a^2}{6}\frac{\partial^2 \psi}{\partial z^2}
+\beta w_0 \delta(z-z_0) \psi(z) = - \varepsilon  \psi(z).
\end{equation}
For very long chains ($N\gg 1$), the sum (\ref{expansion})
is dominated by the largest term, corresponding to the lowest
eigenvalue $(-\varepsilon_0)$ (ground-state dominance):
\begin{equation}
\label{gs_solution} G(z,z';N)
\simeq e^{-\varepsilon_0 N}\psi_0(z)\psi_0^*(z').
\end{equation}

The analogous problem of quantum-mechanical particle motion in the 
potential $V(z)$ has been treated in detail by Aslangul~\cite{Aslangul}. 
To establish the connection to these calculations, we rewrite the 
stationary differential equation (\ref{stat_diff_eq}) in the form
\begin{equation}
\label{stat_diff_eq_par} \psi''(z)-k^2\psi(z) =
- k_0\delta(z-z_0)\psi(z),
\end{equation}
where $k$ and $k_0$ are given by
\begin{equation}
\label{k_def} k^2=-\frac{6\varepsilon_0}{a^2}
\end{equation}
and
\begin{equation}
\label{k0_def} k_0=\frac{6\beta w_0}{a^2}=\frac{6\beta
(\xi_0+v_0)}{a^2}.
\end{equation}
The boundary conditions are: $\psi(z)|_{z=0}=0$ and
$\psi(z)|_{z \to \infty}=0$.  Eq.~(\ref{stat_diff_eq_par}) has
the same form as Eq.~(3) in Ref. \onlinecite{Aslangul}, except that 
we have chosen $k_0>0$, and we can simply apply the results of 
Ref. \onlinecite{Aslangul}. 
For the ground state eigenfunction we obtain
\begin{equation}
\label{psi2} \psi_0(z)=A\left[e^{-k_b|z-z_0|} - e^{-k_b(z+z_0)}
\right]
\end{equation}
where $k_b$, determining the ground-state energy (\ref{k_def}), is
the root of the transcendental equation
\begin{equation}
\label{eq_kstar} \frac{2k}{1-e^{-2k z_0}}=k_0
\end{equation}
and the normalization constant $A$ is equal to
\begin{equation}
\label{A_def2} A=\left(\frac{k_0/2}{1+(2k_b-k_0)z_0}\right)^{1/2}.
\end{equation}
Eq.~(\ref{eq_kstar}) always has the trivial solution $k=0$. 
A nontrivial solution exists if the slope of the line $y=2k$ is smaller
than the slope of the curve $y=k_0(1-e^{-2k z_0})$ at $z=0$. This
condition yields the value of the adsorption transition point.
\begin{equation}
\label{trp_h2} 
k_0^{tr}=\frac{1}{z_0}.
\end{equation}
The expression for the ground-state Greens function is 
\begin{equation}
\label{gs_G2}
\begin{split}
G(z,z';N) = 
& 
\: \frac{k_0/2}{1+(2k_b-k_0)z_0} \left[
e^{-k_b|z-z_0|}
- e^{-k_b(z+z_0)} \right] 
\\ & \times 
\left[e^{-k_b|z'-z_0|} - e^{-k_b(z'+z_0)} \right] \: e^{- \epsilon_0 N}
\end{split}.
\end{equation}

Based on these results we can now calculate the averages in
(\ref{first_cumulant}), using the ground-state dominance approximation
(\ref{gs_solution}) for the Greens function.
\begin{eqnarray}
\label{V_av}
\lefteqn{
\overline{V(z(n))} 
} 
\\ && \nonumber
 = \frac{\int_0^N dl \int_0^\infty dx \, dy \,
dz \, G(x,z;l)\delta(z-z_0)G(z,y;N-l)}{\int_0^N dl
\int_0^\infty dx \, dy \, dz \, G(x,z;l)G(z,y;N-l)} 
\\ & & \nonumber
\simeq \int_0^\infty dz |\psi_0(z)|^2 \delta(z-z_0)
= |\psi_0(z_0)|^2 
\\ & & \nonumber
= A^2\left(1-e^{-2k_bz_0}\right)^2 =
A^2\left(\frac{2k_b}{k_0}\right)^2 =
\frac{2k_b^2/k_0}{1+(2k_b-k_0)z_0}
\end{eqnarray}

Since different replicas are independent, the second term of
Eq.~(\ref{first_cumulant}) becomes
\begin{equation}
\label{V2_av_diff}
\begin{split} 
\overline{V(z_\alpha (n))V(z_\gamma (n+k))}\Big|_{\alpha \ne \gamma}
&
= \overline{V(z_\alpha (n))}^{\: 2} 
\\&
\simeq \left[\frac{2k_b^2/k_0}{1+(2k_b-k_0)z_0}\right]^2
\end{split}
\end{equation}
Inserting these averages in Eq.~(\ref{first_cumulant}) and using
the explicit expression (\ref{corrf}) for the correlation function 
$c(k)$, we obtain for the first cumulant
\begin{equation} \label{first_cumulant2} 
\begin{split}
\frac{\beta \Delta F_1}{N} = 
&
\beta^2\Delta^2 \Real \left\{ S(\Gamma) -
\frac{1}{\Gamma}\left[\frac{2k_b^2/k_0}{1+(2k_b-k_0)z_0}\right]^2 
\right\} 
\\&
-  \beta v_0 \frac{2k_b^2/k_0}{1+(2k_b-k_0)z_0}.
\end{split}
\end{equation}
Here
\begin{equation}
\begin{split}
S(\Gamma)  = 
& 
\int_0^N dk \, \exp(-\Gamma k)
\overline{V(z(n))V(z(n+k))} 
\\
 \simeq  
&
\int_0^\infty dk \, \exp(-\Gamma k)
\overline{V(z(n))V(z(n+k))}
\end{split}
\end{equation}
is the Laplace transform of $\overline{V(z(n))V(z(n+k))}\equiv s(k)$
and we have used 
\begin{equation}
\int_0^N dk \, c(k) = 
\frac{\Delta^2}{\Gamma}\left( 1-e^{-\Gamma N}\right)
\stackrel{N \gg 1}{\simeq} \frac{\Delta^2}{\Gamma}.
\end{equation}
Our next task is to calculate $s(k)$ and $S(p)$. 
By definition, we have
\begin{widetext}
\begin{equation}
\label{s_k}
\begin{split}
s(k) & = \frac{\int_0^N \!\! dl \int_0^\infty dx \, dy \, dz  \, dz' \,
G(x,z;l)\delta(z-z_0)G(z,z';k)V(z')G(z',y;N-l-k)}
{\int_0^N \!\! dl \int_0^\infty dx \, dy
\, dz  \, dz'\, G(x,z;l)G(z,z';k)G(z',y;N-l-k)}
\end{split}
\end{equation}
\end{widetext}
We adopt the ground-state dominance approximation for the 
overall chain and for the side subchains in Eq.~(\ref{s_k}). 
This yields for $s(k)$
\begin{equation}
s(k) \simeq e^{k\varepsilon_0}  \int_0^\infty dz \, dz' \,
V(z)\psi_0(z)V(z')\psi_0(z')G(z,z';k)
\end{equation}
and for the Laplace transform 
\begin{eqnarray}
\label{sp_final}
\lefteqn{
S(p)  \simeq 
\int_0^\infty \!\!\!\!\! dz \, dz' \,
V(z)\psi_0(z)V(z')\psi_0(z')\mathcal{G}(z,z';p+|\varepsilon_0|) 
}
\nonumber \\ &  \nonumber
= & \int_0^\infty \!\!\!\!\!\! dz \, dz' \,
\delta(z\!-\!z_0)\psi_0(z)\delta(z'\!-\!z_0)\psi_0(z')
\mathcal{G}(z,z';p+|\varepsilon_0|)
\\ &  \nonumber
= & [\psi_0(z_0)]^2 \mathcal{G}(z_0,z_0;p+|\varepsilon_0|) 
\\&  
= & \frac{2k_b^2/k_0}{1+(2k_b-k_0)z_0}\mathcal{G}(z_0,z_0;p+|\varepsilon_0|)
\end{eqnarray}
where $\mathcal{G}(z,z';p)=\int_0^\infty G(z,z';N) e^{-pN}dN$ is
the Laplace transform of the Greens function $G(z,z';N)$ with
respect to the variable $N$. The function 
$\mathcal{G}(z_0,z_0;p+|\varepsilon_0|)$ is calculated in the appendix.
The result is
\begin{equation}
\label{g1_z0_z0_rep}
\begin{split}
\mathcal{G}(z_0&,z_0;p) =
\\&
\frac{\sinh\left(\sqrt{6p}\,z_0/a\right)}{a\sqrt{p/6} \:
e^{\sqrt{6p}z_0/a}- \beta (\xi_0+v_0)
\sinh\left(\sqrt{6p}\,z_0/a\right)}.
\end{split}
\end{equation}
Inserting this result in Eqs.~(\ref{sp_final}) and then
(\ref{first_cumulant2}), we obtain
\begin{widetext}
\begin{equation}
\label{first_cumulant_final2}
\begin{split}
& \frac{\beta \Delta F_1}{N}= \frac{2k_b^2/k_0}{1+(2k_b-k_0)z_0} \\
& \times \left\{ \beta^2\Delta^2 \Real \left(
\frac{6\sinh\left(\sqrt{6\Gamma/a^2+k_b^2}\,z_0\right)}{a^2\sqrt{6\Gamma/a^2+k_b^2}
\: e^{\sqrt{6\Gamma/a^2+k_b^2}\,z_0}- 6\beta (\xi_0+v_0)
\sinh\left(\sqrt{6\Gamma/a^2+k_b^2}\,z_0\right)} 
-\frac{2k_b^2
/k_0}{1+(2k_b-k_0)z_0}\frac{1}{\Gamma}\right) -\beta v_0 \right\}
\end{split}
\end{equation}
\end{widetext}
Let us investigate the behavior of this expression near the transition
point (\ref{trp_h2}), taking the limit $z_0 \to 0$.
First, we note that in the vicinity of the transition point 
$k_0^{tr}=1/z_0$, the product $k_0 z_0$ can be represented as
$ k_0 z_0=1+\delta z_0 $.
Inserting this expression into Eq.~(\ref{eq_kstar}) and
considering the case $z_0 \ll 1$ it is easy to show that 
$\delta \approx k_b$. Substituting this representation for $k_0 z_0$ 
into Eq.~(\ref{first_cumulant_final2}), we obtain 
\begin{equation}
\label{first_cumulant_expansion2}
\begin{split}
\frac{\beta \Delta F}{N} {\simeq} 2k_b^2 \beta^2 \Delta^2 \Real
\left[ \frac{1}{\Gamma}\left(\sqrt{1+\frac{6\Gamma}{a^2 k_b^2}}-1
\right) \right] & -2\beta v_0 k_b , 
\\&
\mbox{  for  } z_0 \ll 1.
\end{split}
\end{equation}

\subsection{Determination of the Transition Point}
\label{sec:transition}

In the previous sections, we have developed the general theory 
and obtained expressions for the first cumulant $\Delta F_1$,
an approximate expression for the difference between the free energies 
of the original RHP system and the reference hompolymer system. 
Following the idea of the reference system approach, we now
need to adjust the auxiliary parameter $v_0$ such that the free 
energy of the reference system approximates that of the original
system in an optimal way. This implies that at the optimal 
value of $v_0$, the first cumulant $\Delta F_1$ should be equal to zero:
\begin{equation}
\label{first_cumulant_zero} \Delta F_1(v_0^{tr})=0
\end{equation}
%
Looking at the explicit expression for $\Delta F_1$, 
Eq.~(\ref{first_cumulant_final2}), we note that this
equation always has one trivial solution $k_b=0$ or
$w_0=v_0+\xi_0=0$. This corresponds to the uninteresting case of
a chain which is desorbed both in the original and and in the 
reference system. The other, nontrivial, solution for $v_0$ is obtained 
from the following equation:
\begin{widetext}
\begin{equation}
\label{eq_v02}
\begin{split}
\beta^2 \Delta^2  \Real & \left\{
\frac{6\sinh\left(\sqrt{6\Gamma/a^2+k_b^2}\,z_0\right)}{a^2\sqrt{6\Gamma/a^2+k_b^2}
\: e^{\sqrt{6\Gamma/a^2+k_b^2}\,z_0}- 6\beta (\xi_0+v_0)
\sinh\left(\sqrt{6\Gamma/a^2+k_b^2}\,z_0\right)} 
-\frac{2k_b^2
/k_0}{1+(2k_b-k_0)z_0}\frac{1}{\Gamma}\right\} -\beta v_0 = 0.
\end{split}
\end{equation}
\end{widetext}
The transition point for the reference system corresponds to the
condition $z_0 k_0^{tr}=1$ (\ref{trp_h2}) or, according to the
definition (\ref{k0_def}) of $k_0$, to
\begin{equation}
\label{trp2}
\frac{6\beta(\xi_0^{tr}+v_0)}{a^2}=\frac{1}{z_0}
\end{equation}
Substituting this condition into Eq.~(\ref{eq_v02}) yields an
equation for the transition point $\xi_0^{tr}$. 
It will be discussed in the next section.
\subsection{Final Result}
\label{sec:result_ana}
For the sake of convenience, we introduce the dimensionless variables
\begin{equation}
\label{dimless} 
\widetilde{\xi_0} = \frac{\beta \xi_0}{a}, \quad
\widetilde{\Delta} = \frac{\beta \Delta}{a}, \mbox{ and} \quad
\widetilde{z_0} = \frac{z_0}{a}.
\end{equation}
Our final result for the adsorption transition point 
$\widetilde{\xi_0}^{tr}$ for heteropolymers
(Eqs.~(\ref{eq_v02}) and (\ref{trp2})) can then be cast into the form
\begin{equation}
\label{trp_final2}
\widetilde{\xi_0}^{tr}-\widetilde{\xi_H}^{tr} =
-\widetilde{\Delta}^2 
\Real \left\{
\frac{6\sinh\left(\sqrt{6\Gamma}\,\widetilde{z_0}\right)}{\sqrt{6\Gamma} \:
e^{\sqrt{6\Gamma}\,\widetilde{z_0}} - 1/\widetilde{z_0}
\sinh\left(\sqrt{6\Gamma}\,\widetilde{z_0}\right)} \right\},
\end{equation}
where $\widetilde{\xi_H}^{tr}$ is the transition point for homopolymers.
The latter depends strongly on the specific shape of the surface 
potential and in particular on its range $\widetilde{z_0}$
($\widetilde{\xi_H}^{tr} = 1/6 \widetilde{z_0}$, cf. Eq.~(\ref{trp_h2})).
However, the {\em shift} of the transition point for RHPs relative to
a homopolymer system, Eq.~(\ref{trp_final2}), is much less sensitive 
to variations of $\widetilde{z_0}$. In the limit $\widetilde{z_0} \ll 1$ 
the solution takes the elegant form
\begin{equation}
\label{trp_final_expansion2}
\widetilde{\xi_0}^{tr}-\widetilde{\xi_H}^{tr}  =
-\widetilde{\Delta}^2 \Real \sqrt{\frac{6}{\Gamma}}.
\end{equation}
   
\begin{figure}[t]
\begin{center}
\includegraphics[width=10cm,angle =0]{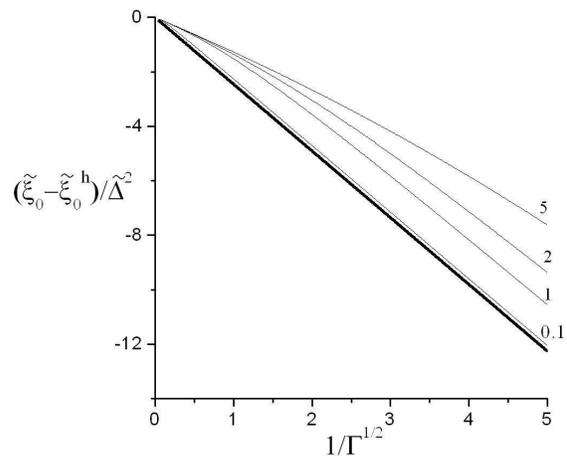}
\end{center}
\caption{
Shift $(\widetilde{\xi}_0^{tr} - \widetilde{\xi}_H^{tr})/\widetilde{\Delta}^2$
of the adsorption-desorption transition point for RHPs 
with exponentially decaying sequence correlations compared to that
of homopolymers, vs. correlation parameter $1/\Gamma^{1/2}$,
according to Eq.~(\ref{trp_final2}). The different curves show
the results for different values of $\widetilde{z_0}$ as
indicated.
\label{fig_exp} 
}
\end{figure}

\begin{figure}[t]
\begin{center}
\includegraphics[width=10cm,angle =0]{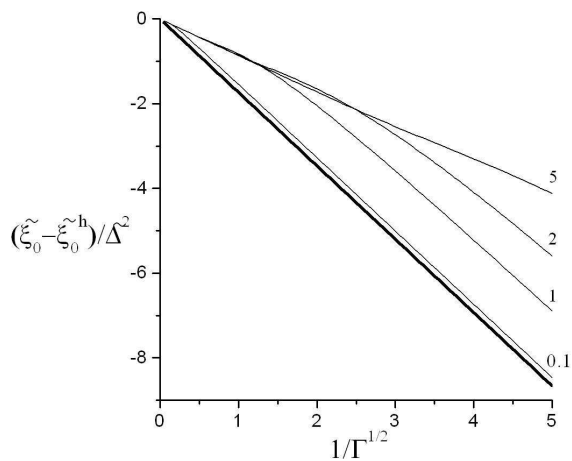}
\end{center}
\caption{
Same as Fig.~\ref{fig_exp}, but for oscillating sequence correlations.
\label{fig_osc} 
}
\end{figure}

Figs.~\ref{fig_exp} and \ref{fig_osc} show adsorption-desorption 
transition curves as a function of the sequence correlation parameter 
$1/\sqrt{\Gamma}$, as calculated from Eq.~(\ref{trp_final2}) for 
different values of $\widetilde{z_0}$. With increasing $1/\sqrt{\Gamma}$,
the curves become straight lines with the same slope as the 
asymptotic curve, Eq.~(\ref{trp_final_expansion2}). 
Only in an interval close to the origin does the slope of the curves 
deviate from the asymptotic value. At small $z_0$ ($\widetilde{z_0} < 1$), 
the simplified expression (\ref{trp_final_expansion2}) provides a
good approximation for the full solution (\ref{trp_final2}).

The expressions (\ref{trp_final2}) and (\ref{trp_final_expansion2}) 
demonstrate that the mean affinity parameter ($\widetilde{\xi_0}$) corresponding
to the adsorption transition is reduced for RHPs, compared to homopolymers. 
In other words, RHPs have a better effective affinity to surfaces
than homopolymers with the same mean affinity.

Our result also shows that the surface affinity of RHPs with exponentially 
decaying correlations is higher than that of RHPs with oscillating
correlations. This can be seen by rewriting the general expression
(\ref{trp_final_expansion2}) for the particular cases of purely 
exponential decay ($\Gamma \in \mathbb{R}$)
\begin{equation}
\label{trp_final1_exp}
\widetilde{\xi_0}^{tr}-\widetilde{\xi_H}^{tr} 
=-{\widetilde{\Delta}}^2 \sqrt{\frac{6}{\Gamma}} \: , 
\: \Gamma \in \mathbb{R}
\end{equation}
and purely oscillating correlations
($\Gamma \in i \mathbb{R}$)
\begin{equation}
\label{trp_final1_osc}
\widetilde{\xi_0}^{tr}-\widetilde{\xi_H}^{tr} 
=-{\widetilde{\Delta}}^2 \sqrt{\frac{3}{\Gamma'}} \: , \Gamma = i\,
\Gamma',\:  \Gamma' \in \mathbb{R}.
\end{equation}

\section{Numerical Approach}
\label{sec:num}
In the previous section, we have developed an analytical theory 
of RHP adsorption based on the reference system approach, and 
subobtained as a main result the adsorption-desorption phase
diagram, Eqs.~(\ref{trp_final2}) and (\ref{trp_final_expansion2})
and Fig.~\ref{fig_exp}. 
To assess directly the validity of the reference system approach, 
and to test our final result, Eq.~(\ref{trp_final_expansion2}),
we have also performed numerical calculations based on a simple 
lattice model, following the method of Scheutjens and 
Fleer \cite{PolInterfaces}.

\subsection{Lattice Model}
\label{sec:model_num}

In the model we use for the numerical calculations, polymer chains
of $N$ monomers are represented by random walks of length $N$ on the 
simple cubic lattice. We consider polymers which have one
end tethered to an impenetrable planar surface.
In the $z$-direction perpendicular to the plane, the center of mass 
of a monomer can have coordinates $z=0, l, 2l, 3l \ldots$, where 
the layer $z=0$ is directly adjacent to the adsorbing plane. 
The adsorption potential is taken to be short-ranged and
sequence-dependent and has the following form
\begin{equation}
\label{l_potential} \frac{u_i\left( z, i \right)}{kT}=\left\{
\begin{array}{ll}
\chi_{M_i} & \mbox{if }z=0 \\
0 & \mbox{if }z>0
\end{array},
\right.
\end{equation}
i.~e., it acts only on the monomers in the layer that is adjacent
to the adsorbing surface.  In Eq.~(\ref{l_potential}) $M_i$
denotes the type of the $i$-th monomer in the chain and $\chi_M$
is the adsorption interaction parameter for monomers of the
type $M$. Note that we consider the simplest case of a phantom
chain, hence the monomers do not interact with each other and the
local monomer potential $u_i\left( z, i\right)$ does not
depend on the local polymer concentration.

One immediately notices one important difference between this system 
and the model considered in the previous section: Up to now we have 
considered \emph{free} chains, whereas we now study \emph{tethered} chains. 
However, within the ground state dominance approximation used to treat 
chain ends in the last section, the final result for the adsorption 
transition will be identical for tethered chains and for free chains. 
Tethering the chains in the numerical calculations
helps to avoid an uncertainty with the normalization, because for 
free chains of finite length one would have to introduce a box of 
finite size.

We consider two-letter RHP sequences constructed by first order Markov chains
($M_s=A, B$). The underlying Markov process is
determined by the probabilities to find single A and B monomers
($f_A$ and $f_B=1-f_A$, respectively), and by the nearest-neighbor
transition probabilities $p_{i\rightarrow j}$ which is the
probability that a monomer $i$ is followed by a monomer $j$. 
It is convenient to introduce the correlation parameter
\begin{equation}
\label{clusterp_def} C = 1-p_{A\rightarrow B}-p_{B\rightarrow A}
\end{equation}
that characterizes the correlations in the sequence. In the
case $C =0$ one has polymers with uncorrelated monomer sequences 
(Bernoullian type), at $C >0$ the probability to encounter nearest-neighbors 
of the same type is enhanced, and at $C <0$ nearest-neighbor monomers are 
more likely to be of different type (as in alternating copolymers). 
The statistical distribution of the sequences is completely determined 
by the two parameters $f_A$ and $C$.

In the actual calculations, A monomers were always considered as 
adsorbing or sticky ($\chi_A$ is positive), whereas monomers of the 
B type were taken to be either neutral ($\chi_B$ = 0) or repelling 
from the surface ($\chi_B=-\chi_A$). These two cases will be
referred to as SN (sticker-neutral) or SR (sticker-repulsive).

\subsection{Numerical Method}
\label{sec:greens}
The statistical weight $G_t(z;N)$ of all conformations of tethered
chains with one free end in the layer $z$ satisfies the following
recurrent relation first introduced by Rubin \cite{Rubin} and 
later used in the more general theory of Scheutjens and Fleer
\cite{PolInterfaces}
\begin{equation}
\begin{split}
\label{recursion} G_t(z;N+1) = &\left\{ \lambda G_t(z-1;N) +
(1-2\lambda)G_t(z;N) 
\right. \\& \left.
+ \: \lambda G_t(z+1;N) \right\}
\end{split}
\end{equation}
\begin{equation}
\label{recursion0} G_t(0;N+1) = \exp(\chi_{M_i}) \left\{
(1-2\lambda)G_t(0;N) + \lambda G_t(1;N) \right\},
\end{equation}
where $\lambda$ is the probability that a random walk step 
connects neighboring layers. On simple cubic lattices, one has 
$\lambda=\frac{1}{6}$. Using as the starting point the monomer 
segment distribution $G(0;1) = \exp(\chi_{M_i})$ and recursively 
applying Eq.~(\ref{recursion}), one can easily calculate 
$G_t(z;N)$ for every chain length $N$.
The statistical weight of all conformations of tethered chains is
then obtained by summing over all positions of the free end:
\begin{equation}
\label{Z_teth} Z=\sum_{z=0}^{Nl} G(z;N)
\end{equation}

The change in the free energy 
of the tethered chain with respect to the free chain in the solution 
is given by
$ \Delta F=-\log Z $.
Here the translational entropy of the free chain has been disregarded,
i.~e., the chains are assumed to be sufficiently long that it can be neglected.
At the transition point one has $\Delta F=0$, i.~e., the energetic benefit
of monomer-surface contacts is equal to the entropic penalty connected 
with the fact that tethering the chain at the plane restricts the number 
of conformations available to the chain.

In order to calculate conformational characteristics of the adsorbed chain, 
one needs to evaluate not only $G_t(z;n)$ for arbitrary $n=1,2,...,N$, 
but also a second set of functions $G_f(z;N-n)$, which gives the statistical 
weight of chain parts between the $n$th monomer and the end monomer $N$, 
subject to the constraint that the monomer $n$ is fixed in the layer $z$ 
(whereas the end monomer is free). 
To calculate $G_f(z;N-n)$ one uses the same recurrence
relations as in Eqs.~(\ref{recursion}) and (\ref{recursion0}), 
but with \emph{reverted} copolymer sequence and different initial 
conditions $G(z;1) = e^{\left. -u_N\left( z\right)\right/kT}$. 
Combining $G_t$ and $G_f$ one can then calculate,
for example, the average fraction of A contacts with the surface
\begin{equation}
\label{s_A} s_A=\frac{1}{\mathcal{N}}\sum_{i=1}^{N} G_t(0;i)
G_f(0;N-i+1) \exp(\chi_{M_i}) \delta_{M_i,A}.
\end{equation}
Here $\delta_{M_i,A}=1$ if $M_i=A$ or 0 otherwise, and the exponential
factor is used to correct for the double contribution of the $i$-th monomer.
The normalization constant $\mathcal{N}$ is given by
\begin{equation}
\label{norm} \mathcal{N}=\sum_{z=0}^{N-1} \sum_{i=1}^{N} G_t(z;i)
G_f(z;N-i+1) e^{\left. u_s\left( z\right)\right/kT} \delta_{M_s,A}.
\end{equation}
The total fractions of A and B contacts can be obtained via
\begin{equation}
\label{s_tot} s=\frac{1}{\mathcal{N}}\sum_{i=1}{N} G_t(0;i)
G_f(0;N-i+1) e^{\left. u_s\left( 0\right)\right/kT}.
\end{equation}

In this paper, we present results for chains of length $N=500$. 
For every set of model parameters $f_A$ and $C$, the adsorption 
characteristics were calculated as a function of the interaction 
parameter $\chi_A$ for 50 different sequence realizations and then
averaged. This corresponds to a situation with \emph{quenched} sequence
disorder. In contrast, the method developed by van Lent and
Scheutjens \cite{vanLent} describes copolymers with
\emph{annealed} disorder.

A full account of the numerical results will be given elsewhere. 
In the present paper, we are mostly interested in the comparison
with the analytical theory, and in particular, in the two questions 
that are discussed in the next section.
\section{Comparison of the Two Approaches}
\label{sec:comparison}
\subsection{Is the reference system approach valid?}
Introducing the reference homopolymer system in Section \ref{sec:approach},
we have assumed that its conformational and thermodynamic
properties are reasonably close to those of the original RHP
system. Later, the optimal value of variational parameter $v_0$
has been chosen according to the requirement that it minimizes the
free energy difference between the original and the reference systems
(\ref{first_cumulant_zero}). Therefore, the free energies of the
two systems are automatically close. However, one cannot be sure that 
this choice of $v_0$ guarantees good correspondence between 
the conformational characteristics. This shall be tested first.

\begin{figure}[t]
\begin{center}
\includegraphics[width=10cm,angle =0]{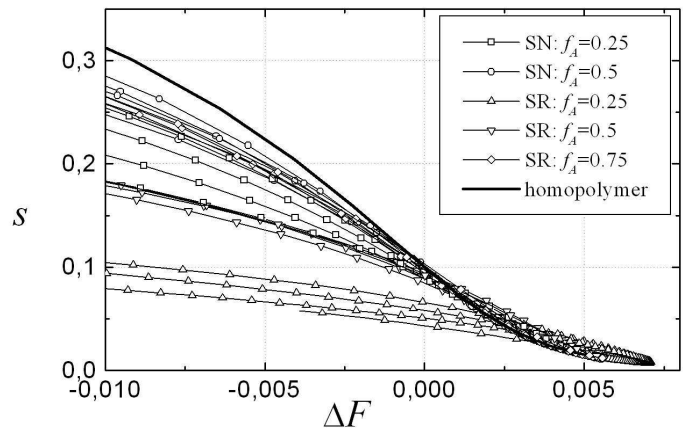}
\end{center}
\caption{
Fraction $s$ of adsorbed segments for homopolymer and RHPs of different
type (SN or SR) and composition ($f_A$) as a function of 
the free energy. For each of RHP type and composition curves 
for $C=0, 0.25, 0.5$ and 0.75 are shown.
\label{fig_s_vs_free} 
}
\end{figure}

One of the most important and indicative conformational property
of the adsorbed chain is undoubtedly the total fraction $s$ of 
adsorbed monomer units. It is plotted in Fig.~\ref{fig_s_vs_free} 
for RHPs with different composition and structure, and for
homopolymers with the same free energy.
One can see that near the adsorption transition 
($\Delta F=0$), the value of $s$ for RHPs is close to that 
for homopolymers. The ``worst'' RHPs are those
which contain a majority of surface-repelling monomers.
In the strong adsorption regime, however, the values of $s$
differ considerably. 

This observation can be explained by the fact that 
in the vicinity of the transition point, chains have relatively 
few surface contacts and long loops and tails. Such conformations
can be realized equally easily by both homopolymers and RHPs, 
in spite of the "gaps" of neutral or repelling monomers in the RHPs.
Since only few units adsorb to the surface, the system
can avoid entropically and energetically unfavorable contacts of
neutral/repelling units with the surface. In the strong adsorption
regime, it is thermodynamically more favorable for homopolymers 
to have many contacts with the surface. This is not necessarily 
the case for RHPs, because non-adsorbing monomer units still avoid
to get involved in the interactions with the surface.

In the strong adsorption regime, the reference system approach
of section \ref{sec:approach} thus becomes questionable.
However, it seems to be applicable in the vicinity of the
transition, and should therefore be suitable to study
the transition point.

\subsection{Is the analytically predicted phase diagram correct?}

Next we check the scaling relation, Eq.~(\ref{trp_final_expansion2}),
describing the shift of the adsorption transition point of RHPs
relative to homopolymers.

To compare Eq.~(\ref{trp_final_expansion2}) with the numerical
results, we need to do two things: (1) find the transition points 
for lattice RHPs with different composition and "structure"
and (2) "translate" these results into of parameters of the continuum 
model parameter
($\widetilde{\xi_0}$, $\widetilde{\Delta}$, and $\Gamma$).

The transition point $\chi_A^{tr}$ in the lattice model was determined
numerically from the condition $\Delta F=0$ for every single 
realization of a RHP. Then the average over the values $\chi^{tr}$
was performed.

The next step is to find the relation between the statistical
parameters of the continuous model and the discrete lattice model. 
For homopolymer systems this problem has been solved by Gorbunov 
et al.~\cite{mapping}. According to this work, the coefficient connecting
the continuous and the discrete polymerization degree is unity, 
and the length $a$ in the continuum model corresponds to the 
lattice spacing $l$ in the case of a simple cubic lattice.
To find the mapping for $\widetilde{\xi_0}$, $\widetilde{\Delta}$,
and $\Gamma$, we calculate the mean monomer-surface interaction
parameter
\begin{equation}
\label{chi_0} \chi_0 = f_A \chi_A + (1-f_A)\chi_B
\end{equation}
and the monomer-monomer correlation function in terms of the 
monomer interaction energy. The calculation is straightforward
and yields
\begin{equation}
\begin{split}
\label{chi_corr} c_\chi(n) \equiv &\langle
\left(\chi(x)-\chi_0\right)\left(\chi(x+n)-\chi_0\right) \rangle
\\
 = & C^n f_A(1-f_A)(\chi_A - \chi_B)^2.
\end{split}
\end{equation}
If we represent the correlation function in an exponential form
like (\ref{corrf})
$$c_\chi(n)=\Delta_\chi^2 \exp\left(-\Gamma_\chi n
\right),
$$
we obtain the equivalent parameters $\Delta$ and $\Gamma$ 
\begin{equation}
\label{Gamma_lat} \Gamma_\chi = -\log(C)
\end{equation}
\begin{equation}
\label{Delta_lat} \Delta_\chi = \left( \chi_A -\chi_B \right)
\sqrt{f_A(1-f_A)}.
\end{equation}
Since the units for the polymerization degree are the same in both
models, this yields $\Gamma = \Gamma_\chi$. 

The two remaining parameters $\widetilde{\xi}_0$ and 
$\widetilde{\Delta}$ are related to the monomer-surface affinity. 
The mapping of this ``energy scale'' is much less straightforward. 
We shall adopt a procedure suggested in Ref. \onlinecite{mapping}
and adjust the energy scale such that the slope of the monomer
density profiles of adsorbed homopolymers at the surface is the
same in the vicinity of the adsorption transition. 
Comparing Eq.~(15) of Ref. \onlinecite{mapping} with
the $z_0 \to 0$ limit of Eq.~({\ref{gs_G2}), we obtain
\begin{equation}
\frac{\widetilde{\xi}_H-\widetilde{\xi}^{tr}_H}{\chi_H-\chi^{tr}_H} 
= \frac{5}{6}
\end{equation}
for homopolymers on simple cubic lattices. We assume that the
factor (5/6) thus relates the energy scales in our two models,
i.~e., the continuum model and the lattice model.
Thus we conclude
\begin{equation}
\label{xi_0_map} 
\frac{\widetilde{\xi}^{tr}_0-\widetilde{\xi}^{tr}_H}{\chi^{tr}_0-\chi^{tr}_H}
=
\frac{\widetilde{\Delta}}{\Delta_\chi}
=
\frac{5}{6}.
\end{equation} 
The approach has the drawback of being rather indirect. 
Moreover, the energy mapping applies, if at all, only in the close 
vicinity of the transition point.

Inserting these relations for $\widetilde{\xi}_0$, $\Gamma$,
and $\widetilde{\Delta}$ into Eq.~(\ref{trp_final_expansion2}) we
obtain the following relation between equivalent lattice
parameters
\begin{equation}
\label{trp_final2_lat}
\chi_0^{tr} - \chi^{tr}_H=-\Delta_\chi^2 \frac{5}{\sqrt{6\Gamma}}
\end{equation}
Thus $\chi_0^{tr} - \chi^{tr}$ is expected to depend
linearly on $\Delta_\chi^2/\sqrt{\Gamma}$, with a slope 
$-5/\sqrt{6}\approx -2.04$.

The results for $\chi_0^{tr}-\chi^{tr}_H$ are shown in Fig.~\ref{fig_scaling}. 
One can see that the numerical calculations reproduce well the
expected linear dependence between $\chi_0^{tr}-\chi^{tr}_H$
and $\Delta_\chi^2 / \sqrt{\Gamma}$. However, the slope 
($\approx -0.754$) is different from that predicted by 
Eq.~(\ref{trp_final2_lat}).

\begin{figure}[t]
\begin{center}
\includegraphics[width=10cm,angle =0]{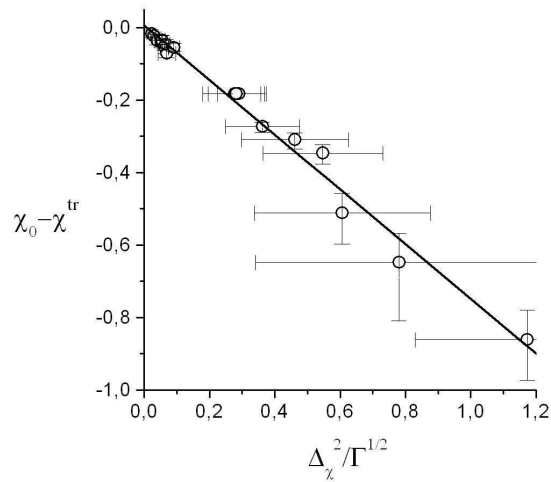}
\end{center}
\caption{
Shift $\chi^{tr}_0-\chi_H^{tr}$ of the adsorption transition in RHP systems 
relative to homopolymer systems for the lattice model 
vs. $\Delta_\chi^2 / \Gamma^{1/2}$.
The solid line shows the best linear fit.
\label{fig_scaling} 
}
\end{figure}

\section{Conclusions}
\label{sec:conclusions}
To summarize, we have studied the adsorption of 
single ideal heteropolymer chains onto homogeneous planar
surfaces. After applying the replica trick we have introduced a
reference homopolymer system chosen such that the
thermodynamic properties were as close as possible to those
of the original RHP system.
This approach allowed us to treat the problem analytically and
to obtain the adsorption-desorption phase diagram of the RHP. 
In particular, we have considered the case of RHPs with 
primary sequence correlations described by a de-Gennes correlator.
We found that the adsorption is enhanced with increasing strength 
of correlation ($\Delta$) and correlation length ($1/\Gamma$), 
both for the cases of decaying and oscillating correlations. 
Our result could be summarized in a very simple equation 
for the shift of the adsorption transition in correlated RHPs 
relative to homopolymers, Eq.~(\ref{trp_final_expansion2}).

To test the reference system approach, numerical lattice
calculations were performed. These results show that in the
vicinity of the transition point, our variational is well
justified. Furthermore, the calculation of the adsorption-desorption 
phase diagram confirmed the analytical scaling prediction.
However, we also found that far from the transition point, one cannot 
"construct" a reference homopolymer system that is good in both
a conformational and a thermodynamical sense. A reference system
perhaps more adequate to describe this case could be a
\emph{multiblock}-copolymer system with block lengths equal to the 
average length of A and B blocks in the random copolymers. 
In the lattice model, these are equal to $s_A=1/[(1-f_A)(1-c)]$ and 
$s_B=1/[f_A(1-c)]$, respectively \cite{rcp_stat}). 
This could be interesting for future work.

We found that our main result, Eq.~(\ref{trp_final_expansion2}),
describes the numerical result very satisfactorily from a 
qualitative point of view. Unfortunately, true quantitative
agreement could not yet be established. The reason lies partly
in the fact that the identification of energy scales in the
lattice model and in the continuum model is not self evident.
We have tested a mapping procedure which adjusts the monomer 
profiles for adsorbed homopolymers. Within this approach,
the theory seems to overestimate the shift of the adsorption 
transition in RHP systems. However, the mapping can be
questioned, as we have discussed in the last section.
Furthermore, we are comparing a theory for RHPs with 
{\em continuous} (Gaussian distributed) monomers with 
numerical calculations for two-letter RHPs, which might also 
lead to problems. 

\section*{Acknowlegdement}
\label{Acknowlegdement}

The financial support of the Deutsche Forschungsgemeinschaft 
(SFB 613) is gratefully acknowledged.

\begin{widetext}

\section*{Appendix: Calculation of $\mathcal{G}(z_0, z_0; p)$}

Our starting point is Eq.~(\ref{diff_eq})
with the initial condition (\ref{initial_cond}) and the boundary
conditions (\ref{boundary_cond}).
Laplace transforming with respect to $N$ gives
\begin{equation}
-p\mathcal{G}(z,z';p)+G(z,z';0)=-\frac{a^2}{6}\frac{\partial^2
\mathcal{G}}{\partial z^2}-\beta w_0 \delta(z-z_0)
\mathcal{G}(z,z';p),
\end{equation}
where $\mathcal{G}(z,z';p)$ is the Laplace transform of
$G(z,z';N)$ with respect to $N,$
$\mathcal{G}(z,z';p)=\int_0^\infty G(z,z';N) e^{-pN}dN$. Taking
into account the initial condition (\ref{initial_cond}), we obtain
\begin{equation}
\label{Lapl1}
-p\mathcal{G}(z,z';p)+\delta(z-z')=-\frac{a^2}{6}\frac{\partial^2
\mathcal{G}}{\partial z^2}-\beta w_0 \delta(z-z_0)
\mathcal{G}(z,z';p)
\end{equation}
with the boundary conditions
\begin{equation}
\label{Lapl1_bc}
\mathcal{G}(z,z';p)|_{z \to \infty}=0, \quad
\mathcal{G}(z,z';p)|_{z \to 0}=0
\end{equation}
Laplace transforming (\ref{Lapl1}) with respect to $z$ with
(\ref{boundary_cond}) gives
\begin{equation*}
-p g(s,z';p)+e^{-z's}= 
-\frac{a^2}{6}\left[
s^2g(s,z';p)-s\mathcal{G}(0,z';p)-\mathcal{G}'(0,z';p)\right] 
-\beta w_0 \mathcal{G}(z_0,z';p)e^{-z_0 s}
\end{equation*}
where $g(s,z';p)$ is the Laplace transform of
$\mathcal{G}(z,z';p)$ with respect to $z$,
$g(s,z';p)=\int_0^\infty \mathcal{G}(z,z';p) e^{-sz}dz$,
and $\mathcal{G}'$ stands
for $\partial \mathcal{G}/ \partial z$. With the
boundary condition at $z=0$ (\ref{Lapl1_bc}) we obtain
\begin{equation}
\label{Lapl2}
-p g(s,z';p)+e^{-z's}= 
-\frac{a^2}{6}\left[
s^2g(s,z';p)-\mathcal{G}'(0,z';p)\right] 
-\beta w_0 \mathcal{G}(z_0,z';p)e^{-z_0 s}
\end{equation}
Solving this equation for $g(s,z';p)$ yields
\begin{equation}
\label{g2}
g(s,z';p)=\frac{\mathcal{G}'(0,z';p)-(6\beta w_0/a^2)
\mathcal{G}(z_0,z';p)e^{-z_0 s}-(6/a^2)e^{-z' s}}{s^2-(6p/a^2)}
\end{equation}
Taking the inverse Laplace transform of (\ref{g2}) with
respect to $s$ gives
\begin{equation}
\label{g1}
\begin{split}
\mathcal{G}(z,z';p)=
\frac{a}{\sqrt{6p}} & \left\{ 
\sinh\left(\sqrt{6p}\,\frac{z}{a}\right) \, \mathcal{G}'(0,z';p) \right. 
-\frac{6\beta w_0}{a^2} \,
\sinh\left(\sqrt{6p}\,\frac{z-z_0}{a}\right) \, 
\mathcal{G}(z_0,z';p) \, \theta(z-z_0)\\
& \left. -\frac{6}{a^2} \sinh \left( \sqrt{6p} \, \frac{z-z'}{a}
\right) \, \theta(z-z') \right\},
\end{split}
\end{equation}
where $\theta(z)$ is the Heaviside function
\begin{equation*}
\theta(z)=\left\{
\begin{aligned}
& 1 \mbox{, if } z>0 \\
& 0 \mbox{, if } z<0
\end{aligned}
\right..
\end{equation*}
Setting $z=z_0$ in Eq.~(\ref{g1}), we obtain 
\begin{equation}
\label{g1_z0}
\mathcal{G}(z_0,z';p)= \, \frac{a}{\sqrt{6p}} 
\left\{\sinh\left(\sqrt{6p}\,\frac{z_0}{a}\right)
\, \mathcal{G}'(0,z';p) 
-\frac{6}{a^2}
\sinh\left(\sqrt{6p}\,\frac{z_0-z'}{a}\right) \,
\theta(z_0-z')\right\}
\end{equation}
and then insert $\mathcal{G}(z_0,z';p)$ back into Eq.~(\ref{g1})
(second term):
\begin{equation}
\label{g1_semifinal}
\begin{split}
\mathcal{G}(z,z';p)
& = 
\frac{a}{\sqrt{6p}} 
\left\{ \sinh\left(\sqrt{6p}\,\frac{z}{a}\right) \, 
\mathcal{G}'(0,z';p)-\frac{6\beta w_0}{a^2}
\, \sinh\left(\sqrt{6p}\,\frac{z-z_0}{a}\right) \, \theta(z-z_0) \right.\\
& \times \frac{a}{\sqrt{6p}}
\left[\sinh\left(\sqrt{6p}\,\frac{z_0}{a}\right) \,
\mathcal{G}'(0,z';p) -\frac{6}{a^2}
\sinh\left(\sqrt{6p}\,\frac{z_0-z'}{a}\right) \, \theta(z_0-z')
\right]\\
& \left. -\frac{6}{a^2}
\sinh\left(\sqrt{6p}\,\frac{z-z'}{a}\right) \, \theta(z-z')
\right\}
\end{split}
\end{equation}
To find the unknown $\mathcal{G}'(0,z';p)$, we require
$\mathcal{G}(z \to \infty ,z';p)$ to vanish and $z'$ to be finite. 
This implies that the coefficient of the growing term
term $e^{\sqrt{6p}z/a}$ must be zero. 
Using (\ref{g1_semifinal}), this requirement
leads to the following result for $\mathcal{G}'(0,z';p)$:
\begin{equation}
\label{g1_prime_0}
\mathcal{G}'(0,z';p)=\frac{6}{a^2} \: \frac{{e^{-\sqrt{6p}z'/a} -
\frac{6\beta
w_0}{a^2}\frac{a}{\sqrt{6p}}\sinh\left(\sqrt{6p}\,
\frac{z_0-z'}{a}\right)e^{-\sqrt{6p}z_0/a}
\theta(z_0-z')}}{1- \frac{6\beta
w_0}{a^2}\frac{a}{\sqrt{6p}}
\sinh\left(\sqrt{6p}\,\frac{z_0}{a}\right)e^{-\sqrt{6p}z_0/a}}
\end{equation}
To find $\mathcal{G}(z_0,z_0;p)$, we substitute (\ref{g1_prime_0})
into (\ref{g1_semifinal}) and set $z=z_0$ and $z'=z_0$. The result
is
\begin{equation}
\label{g1_z0_z0}
\begin{split}
\mathcal{G}(z_0,z_0;p) & = \frac{a}{\sqrt{6p}} \: \frac{6}{a^2} \:
\frac{\sinh\left(\sqrt{6p}\,z_0/a\right)}{e^{\sqrt{6p}z_0/a}-
\frac{6\beta w_0}{a^2} \frac{a}{\sqrt{6p}}
\sinh\left(\sqrt{6p}\,z_0/a\right)} \\
& = \frac{\sinh\left(\sqrt{6p}\,z_0/a\right)}{a\sqrt{p/6} \:
e^{\sqrt{6p}z_0/a}- \beta w_0 \sinh\left(\sqrt{6p}\,z_0/a\right)}
\end{split}.
\end{equation}
\end{widetext}


\end{document}